\documentclass[aps,prb,reprint,nofootinbib]{revtex4-2}
\usepackage{amsmath,amssymb,amsfonts}
\usepackage{booktabs}
\usepackage{graphicx}
\usepackage{hyperref}
\hypersetup{colorlinks=true, linkcolor=blue, citecolor=blue, urlcolor=blue}

\newcommand{\sx}{\sigma^x}
\newcommand{\sy}{\sigma^y}
\newcommand{\sz}{\sigma^z}
\newcommand{\Dz}{\Delta_z}
\newcommand{\Mz}{\bar{M}_z}
\newcommand{\sn}{\operatorname{sn}}
\newcommand{\cn}{\operatorname{cn}}
\newcommand{\dn}{\operatorname{dn}}
\newcommand{\scn}{\operatorname{sc}}
\newcommand{\arccosh}{\operatorname{arccosh}}

\begin{document}

\title{XY-anisotropy dependence of the spontaneous staggered magnetization
of the gapped XYZ spin-$1/2$ chain}

\author{Felipe Taha Sant'Ana}
\affiliation{Center of Mathematics, Computing, and Cognition, Federal University of ABC, 09210-580, Santo Andr\'e, Brazil}
\email{t.felipe@ufabc.edu.br}

\date{\today}

\begin{abstract}
We determine the spontaneous staggered magnetization of the gapped XYZ spin-1/2 chain as a function of its Hamiltonian couplings. The eight-vertex solution gives this order parameter in elliptic variables, which depend on the couplings only implicitly. We invert the relation between the two sets of variables algebraically, which turns the Baxter--Kelland--Jimbo--Miwa--Nakayashiki polarization formula into an explicit function of the XY and longitudinal anisotropies over the whole antiferromagnetic phase. Expanding this function at small XY anisotropy and fixed longitudinal anisotropy, we obtain the coefficient of the leading quadratic correction in closed form and prove that it is negative. Modular transformations of the elliptic products then give the behaviour of the coefficient near the isotropic point, including a joint scaling limit in the two anisotropies, and the vanishing of the magnetization at the boundary of the phase. We check the closed form against a fourth-order perturbative calculation in the spin chain, independent of the eight-vertex solution, and against infinite-DMRG simulations.
\end{abstract}

\maketitle

\section{Introduction}\label{sec:intro}

The XYZ spin-$1/2$ chain, with three independent nearest-neighbour exchange couplings $J_x$, $J_y$, $J_z$ on a one-dimensional lattice, is the spin-chain limit of the Baxter eight-vertex model~\cite{Baxter1972, Baxter1973I, Baxter1973II, Baxter1973III} whose solution and vertex-operator reformulation are standard~\cite{BaxterBook, JimboMiwaBook}. In this work we parametrize the model by a uniform exchange scale $J>0$, a lattice $XY$ anisotropy $\gamma$ controlling the imbalance between the $\sx\sx$ and $\sy\sy$ couplings, and a longitudinal anisotropy $\Dz$ on the $\sz\sz$ bond, in standard Pauli normalization; the Hamiltonian is given in Sec.~\ref{sec:result}. Several quantities derived from that solution are nonetheless not available in convenient closed form, primarily because the elliptic parametrization used in the integrability framework is not directly tied to the Hamiltonian variables $(\gamma, \Dz)$. The present work treats one such quantity: the leading correction to the staggered order parameter under variation of $\gamma$ at fixed $\Dz$.

The gapped N\'eel regime is realized in Ising-like chain compounds such as SrCo$_2$V$_2$O$_8$, where Bethe-ansatz predictions for strongly anisotropic spin-$1/2$ chains have been tested quantitatively~\cite{Wang2018}. Rydberg-atom arrays engineer XXZ Hamiltonians with tunable couplings~\cite{Scholl2022}, and Floquet driving of a Rydberg gas has produced XYZ ones~\cite{Geier2021}.

In the antiferromagnetic regime $\Dz>1$, for $|\gamma|<\Dz-1$, the residual Ising $\mathbb{Z}_2$ symmetry of the Hamiltonian is spontaneously broken into two pure N\'eel-ordered ground states. The spontaneous staggered magnetization, defined in the N\'eel-up sector as the long-distance limit of the alternating one-point function of $\sz$, is strictly positive. At $\gamma=0$ its value is given by Baxter's classical XXZ formula~\cite{BaxterBook} as an infinite product of hyperbolic tangents of integer multiples of $\arccosh\Dz$; this formula was later given a rigorous finite-$N$ determinant representation~\cite{IKMT1999}. The extension to $\gamma\neq 0$ is given implicitly by the Baxter--Kelland conjecture~\cite{BaxterKelland1974} for the spontaneous polarization of the eight-vertex model, written in a compact $q$-Pochhammer form and proved in Ref.~\cite{JMN1993}. Free-field / screening representations~\cite{LashkevichPugai1998, Lashkevich2002} reproduce that result in the vertex-operator language and provide a general scheme for eight-vertex correlation functions and form factors. Closely related form-factor methods underlies the correlation-amplitude results for the XXZ chain of Refs.~\cite{Lukyanov1999, LukyanovTerras2003}.

The entanglement entropies of the chain have been obtained in closed form~\cite{Ercolessi2010, Ercolessi2011}, in a calculation that likewise turns on translating elliptic into Hamiltonian variables. On the supersymmetric line, combinatorial and Painlev\'e-VI methods give finite-$N$ correlation functions~\cite{RazumovStroganov2010, HagendorfFendley2012, HagendorfRosengren2024, BazhanovMangazeev2010, BazhanovMangazeev2007, Zinn2013}, and simple expressions for the order parameters along that line, in the couplings of the chain, were conjectured in Ref.~\cite{FendleyHagendorf2010} and compared there with the eight-vertex polarization. Analytic limits connect the chain to sine-Gordon~\cite{LukyanovZamolodchikov1997} and to the massless XXZ chain~\cite{CauxKonnoSorrellWeston2012, Lukyanov1999}. The explicit small-$\gamma$ expansion of the spontaneous magnetization in the Hamiltonian variables $(\gamma, \Dz)$ has, to our knowledge, not been recorded in any of these works. The translation between the elliptic parameters $(\lambda, k)$ and the Hamiltonian variables $(\gamma, \Dz)$ involves an implicit-function step: at fixed physical $\Dz$ the crossing parameter $\lambda$ shifts with $\gamma$, and the auxiliary variable $\pi\lambda/[2K(k)]$ of the Jimbo--Miwa--Nakayashiki (JMN) polarization formula moves further through the $\gamma$-dependence of the complete elliptic integral $K(k)$. The shift of $\lambda$ dominates the coefficient and fixes its sign; the $K(k)$ factor cancels the $\eta_0$-linear part of that shift and changes the magnitude. The result is a compact formula with a short negativity proof.

The results also give closed-form lattice benchmarks for tensor-network simulations of the gapped XYZ chain. Matrix-product-state methods~\cite{White1992, Schollwoeck2011} determine the order parameters of gapped chains to high precision, and infinite-DMRG and VUMPS~\cite{ZaunerStauber2018} calculations are routine at bond dimension of order one hundred in the moderately gapped regime. The demanding tests lie near the isotropic point and near the boundary of the N\'eel phase, where the correlation length grows; there the near-critical expansion of Sec.~\ref{sec:asymp} and the full $\gamma$ dependence of Sec.~\ref{sec:full} are the targets.

Our main result, stated in Sec.~\ref{sec:result}, is a closed form for the coefficient $C_M(\Dz)$ of the quadratic-in-$\gamma$ correction to the ratio of the perturbed to the unperturbed magnetization: a rational term in $\Dz^2-1$ plus a Lambert series in reciprocal hyperbolic sines of multiples of $\arccosh\Dz$. The same inversion of the coupling relations gives $\Mz$ as an explicit function of $(\gamma,\Dz)$ over the whole N\'eel phase $|\gamma|<\Dz-1$. At its boundary the chain is a critical XXZ chain, and the order parameter vanishes with the exponent $(\pi-\mu)/(4\mu)$ fixed by the anisotropy $\cos\mu$ of that chain. Near the isotropic point the Lambert sum has only two algebraic terms (Appendix~\ref{app:lambert}), so the divergence of $C_M$ there can be expanded to any order, and a uniform estimate of the logarithm of the order-parameter ratio gives the joint scaling limit in $\gamma$ and $\Dz-1$ (Sec.~\ref{sec:asymp}). Infinite-DMRG at four anisotropies between $\Dz = 1.5$ and $5$ is consistent with the formula. The near-critical results are the lattice input for a continuum sine-Gordon analysis of the staggered moment near the gapless Heisenberg point, in the spirit of~\cite{LukyanovZamolodchikov1997, LukyanovTerras2003}, discussed in Sec.~\ref{sec:discussion}.

Section~\ref{sec:derivation} derives the closed form from the JMN polarization formula and identifies the shift that fixes its sign, Sec.~\ref{sec:asymp} treats the asymptotic limits and the joint scaling limit, and Sec.~\ref{sec:full} gives the full $\gamma$ dependence and its behaviour at the phase boundary. Appendices~\ref{app:jacobi} and~\ref{app:lambert} collect the elliptic and modular details; Appendix~\ref{app:JMN-aux} derives the couplings from the eight-vertex weights, keeping the crossing and spectral parameters distinct; and Appendix~\ref{app:pert} rederives the leading strong-coupling behaviour of $C_M$ by fourth-order perturbation theory in the spin chain.

\section{Model and main result}\label{sec:result}

We consider the XYZ spin-$1/2$ Hamiltonian
\begin{align}\label{eq:H}
H(\gamma) = J\sum_n\left[\right.&(1+\gamma)\sx_n\sx_{n+1}
+ (1-\gamma)\sy_n\sy_{n+1} \nonumber\\
&\left.+ \Dz\sz_n\sz_{n+1}\right],
\end{align}
where $\sigma^\alpha_n$ are Pauli matrices (eigenvalues $\pm 1$); in terms of spin-$1/2$ operators $S^\alpha = \sigma^\alpha/2$ the exchange constants are $4J(1\pm\gamma)$ and $4J\Dz$. The relabeling $x \leftrightarrow y$, a global $\pi/2$ rotation about the $z$ axis, maps $\gamma\to-\gamma$ and leaves $\sz_n$ invariant; every observable invariant under this rotation, the staggered magnetization among them, is therefore an even function of $\gamma$. (Observables that are odd under the relabeling, such as the difference of the $xx$ and $yy$ correlators, are odd in $\gamma$.)

\paragraph*{Main result.}
For the XYZ chain~\eqref{eq:H} with $J > 0$ and $\Dz > 1$, and for $|\gamma| < \Dz - 1$, the model is in the antiferromagnetic gapped phase with two symmetry-broken N\'eel-ordered ground states~\cite{BaxterBook, denNijs1981}. In the N\'eel-up sector, the spontaneous staggered magnetization
\begin{equation}\label{eq:Mz_def}
\Mz(\gamma, \Dz) = \lim_{n\to\infty}(-1)^n\langle\sz_n\rangle_{N\uparrow}
\end{equation}
is strictly positive and analytic in $\gamma^2$ at $\gamma = 0$. Here $\langle\cdot\rangle_{N\uparrow}$ denotes the expectation value in the pure infinite-volume N\'eel-up state, selected by an infinitesimal staggered field: one adds $-h\sum_n(-1)^n\sz_n$ to Eq.~\eqref{eq:H}, takes the thermodynamic limit, and sends $h\to 0^+$ at the end. The leading expansion reads
\begin{equation}\label{eq:expansion}
\frac{\Mz(\gamma, \Dz)}{\Mz(0, \Dz)} = 1 + C_M(\Dz)\gamma^2 + O(\gamma^4),
\end{equation}
where
\begin{align}\label{eq:CM}
C_M(\Dz) = \frac{1}{4(\Dz^2-1)^2}
- \frac{3\Dz}{(\Dz^2-1)^{3/2}}\sum_{n=1}^\infty \frac{n}{\sinh(2n\eta_0)},
\end{align}
with $\eta_0 \equiv \arccosh\Dz$. The coefficient is bounded above by $-5/[4(\Dz^2-1)^2]$, hence negative, for all $\Dz > 1$; analyticity in $\gamma^2$ is shown in Sec.~\ref{sec:derivation}.

\begin{figure}[t]
  \centering
\includegraphics[width=\columnwidth]{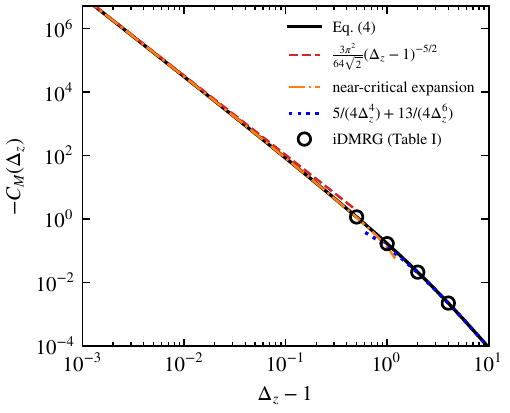}
  \caption{The coefficient $-C_M(\Dz)$ of Eq.~\eqref{eq:CM} across the gapped antiferromagnetic phase $\Dz > 1$ (solid black), together with the leading near-critical asymptote $\frac{3\pi^2}{64\sqrt{2}}(\Dz-1)^{-5/2}$ of Eq.~\eqref{eq:critical} (dashed red), the near-critical expansion of Eq.~\eqref{eq:critical_eta} (dash-dotted orange), and the large-$\Dz$ asymptote $5/(4\Dz^4) + 13/(4\Dz^6)$ of Eq.~\eqref{eq:largeDz_Delta} (dotted blue), shown over their respective regions of validity. Open circles are the iDMRG values of Table~\ref{tab:validation}.}
  \label{fig:CM_overview}
\end{figure}

The first term in~\eqref{eq:CM} comes from the nome dependence of the JMN polarization formula at fixed auxiliary variable and is positive. The second, negative and dominant, term comes from the shift of that auxiliary variable at fixed $\Dz$. The Lambert-type series appearing in~\eqref{eq:CM} has an Eisenstein representation,
\begin{equation}\label{eq:eis_prev}
\sum_{n=1}^\infty\frac{n}{\sinh(2n\eta)} = \frac{E_2(2\tau) - E_2(\tau)}{12}, \qquad \tau = \frac{i\eta}{\pi},
\end{equation}
derived in Appendix~\ref{app:lambert}. Series of this type are classical~\cite{Glaisher1889, Berndt1989} and recur throughout eight-vertex calculations. The bound follows from the $n = 1$ term of the Lambert sum alone (Sec.~\ref{sec:derivation}); it is not optimal at finite $\Dz$ but becomes sharp as $\Dz\to\infty$. As $\Dz \to \infty$, $C_M$ vanishes as $-5/(4\Dz^4)$; as $\Dz \to 1^+$ it diverges as $(\Dz-1)^{-5/2}$, with the subleading terms given in Sec.~\ref{sec:asymp}. In the joint limit $\eta_0\to0^+$ with $\gamma=\omega\eta_0^{5/2}$, uniformly for $\omega$ in compact subsets of $\mathbb{R}$,
\begin{equation}\label{eq:double_scaling_main}
\frac{\Mz(\omega\eta_0^{5/2},\cosh\eta_0)}{\Mz(0,\cosh\eta_0)}
\longrightarrow \exp\left(-\frac{3\pi^2}{16}\omega^2\right).
\end{equation}
Figure~\ref{fig:CM_overview} shows the coefficient with its two asymptotic regimes and the iDMRG values of Sec.~\ref{sec:numerics}. Section~\ref{sec:full} gives $\Mz(\gamma,\Dz)$ over the whole phase and its vanishing at the boundary $|\gamma| = \Dz - 1$.

Writing $H(\gamma) = H(0) + V_\gamma$, the $U(1)$-breaking perturbation is
\begin{equation}\label{eq:Vgamma}
V_\gamma = 2J\gamma\sum_n(\sigma^+_n\sigma^+_{n+1} + \mathrm{H.c.}).
\end{equation}
It changes the total $S^z$ by $\pm2$, and at second order it contributes both diagonal and interference terms to the staggered moment; the sign of their sum follows from the series bound of Sec.~\ref{sec:derivation}. At strong coupling, the Rayleigh--Schr\"odinger calculation of Appendix~\ref{app:pert} evaluates the two contributing terms and gives $C_M=-5/(4\Dz^4)+O(\Dz^{-6})$.

\section{Derivation}\label{sec:derivation}

The Baxter--Kelland spontaneous polarization in the antiferroelectrically ordered regime (regime III)~\cite{BaxterKelland1974, JMN1993} factorizes as $P_0(x, q) = f(q)g(x)$ with
\begin{align}\label{eq:JMN}
f(q) &= \left[\frac{(-q;q)_\infty}{(q;q)_\infty}\right]^2 = \frac{(q^2;q^2)_\infty^2}{(q;q)_\infty^4}, \nonumber\\
g(x) &= \left[\frac{(x^2;x^2)_\infty}{(-x^2;x^2)_\infty}\right]^2,
\end{align}
where $q = e^{-\pi K'/K}$ and $x = \exp[-\pi\lambda/(2K(k))]$ are the elliptic variables, $\lambda$ is the crossing parameter, and $(a;b)_\infty = \prod_{n\geq 0}(1-ab^n)$. Appendix~\ref{app:JMN-aux} fixes these conventions; we write $x = e^{-\eta_x}$ with $\eta_x = \pi\lambda/(2K(k))$. The XXZ limit corresponds to $q = 0$, with $f(0) = 1$ and $g(e^{-\eta}) = \prod_n \tanh^2(n\eta) = \Mz^{\rm XXZ}(\eta)$, recovering Baxter's classical XXZ formula. Taking the logarithm of the first product gives the absolutely convergent series
\begin{equation}
\log f(q)=2\sum_{n=1}^{\infty}[\log(1+q^n)-\log(1-q^n)].
\end{equation}
The $n=1$ term contributes $4q+O(q^3)$, the $n=2$ term contributes $4q^2+O(q^6)$, and all $n\geq3$ terms are $O(q^3)$. Hence
\begin{equation}\label{eq:logf}
\log f(q) = 4q + 4q^2 + O(q^3).
\end{equation}

To translate the elliptic variables into Hamiltonian variables, we work in regime III with $2\eta_B = i\eta$. Jacobi's imaginary transformation, $\sn(i\eta;k) = i\scn(\eta;k')$ and its companions, gives the relations
\begin{align}\label{eq:dict}
\gamma &= k\sinh^2\eta + O(k^3), \qquad q = \frac{k^2}{16} + O(k^4), \nonumber\\
\Dz &= F(\eta, k) \equiv \frac{\dn(\eta;k')}{\cn^2(\eta;k')},
\end{align}
in which the relation for $\Dz$ holds at all $k$, and the first entry is the small-$k$ expansion of the relation $\gamma = k\scn^2(\eta;k')$ (derived in Appendix~\ref{app:JMN-aux} from the eight-vertex weights, including the sublattice rotation connecting~\eqref{eq:H} to the eight-vertex chain). Since $\sn^2(i\eta;k)=-\scn^2(\eta;k')=-\sinh^2\eta+O(k^2)$, with $\scn(\eta;1)=\sinh\eta$ and $k'=\sqrt{1-k^2}$, the relation $\gamma=-k\sn^2(i\eta;k)$ gives $\gamma=k\sinh^2\eta+O(k^3)$. At $k=0$, $\dn(\eta;1)=\cn(\eta;1)=\operatorname{sech}\eta$, so $F(\eta,0)=\cosh\eta$ and $\Dz=\cosh\eta_0$.

The variable $\eta$ can also be eliminated algebraically. With $r=\scn^2(\eta;k')$, the Jacobi identities give
\begin{equation}
\cn^2(\eta;k')=\frac{1}{1+r},\qquad
\dn^2(\eta;k')=\frac{1+k^2r}{1+r}.
\end{equation}
so that the relations for $\gamma$ and $\Dz$ become
\begin{equation}\label{eq:algebraic_dictionary}
\Dz^2=(1+r)(1+k^2r),\qquad \gamma=kr.
\end{equation}
At fixed $\Dz=\cosh\eta_0$, elimination of $r$ gives
\begin{equation}\label{eq:k_equation}
\sinh^2\eta_0=\gamma^2+\gamma\left(k+\frac{1}{k}\right).
\end{equation}
The branch satisfying $k(0)=0$ is
\begin{align}\label{eq:k_exact}
k(\gamma)&=\frac{2\gamma}{\sinh^2\eta_0-\gamma^2+\sqrt{(\sinh^2\eta_0-\gamma^2)^2-4\gamma^2}}\nonumber\\
&=\frac{\gamma}{\sinh^2\eta_0}+\frac{\cosh^2\eta_0}{\sinh^6\eta_0}\gamma^3+O(\gamma^5).
\end{align}
This branch is analytic and odd in $\gamma$ near $\gamma=0$, and
\begin{equation}\label{eq:k2_conversion}
\left.\partial_{\gamma^2}k^2\right|_0=\frac{1}{\sinh^4\eta_0}.
\end{equation}
The complete-integral expansion needed below is
\begin{equation}
K(k)=\frac{\pi}{2}\left(1+\frac{k^2}{4}+O(k^4)\right).
\end{equation}
For the nome, the theta-constant identity gives
\begin{equation}
k^2=\frac{\vartheta_2(0,q)^4}{\vartheta_3(0,q)^4}
=16q-128q^2+O(q^3).
\end{equation}
Analytic inversion at $q=0$ gives
\begin{equation}\label{eq:nome_derivation}
q=e^{-\pi K'/K}=\frac{k^2}{16}+\frac{k^4}{32}+O(k^6).
\end{equation}
The nome is therefore analytic in $k^2$ at the origin.

These relations establish the analyticity stated in Sec.~\ref{sec:result}. Since $\partial_\eta F(\eta_0,0)=\sinh\eta_0\neq0$, the equation $F(\eta,k)=\Dz$ defines an analytic function $\eta(k^2)$ near $k=0$; $K(k)$ and $q$ are analytic in $k^2$; and by Eq.~\eqref{eq:k_exact} $k^2$ is an analytic function of $\gamma^2$. The logarithms of the products in Eq.~\eqref{eq:JMN} converge normally on compact subsets of $|q|<1$, $|x|<1$, by comparison with $\sum_n |q|^n$ and $\sum_n |x|^{2n}$, and in the physical regime $0<q,x<1$ every factor is positive. Hence $P_0>0$, and $\log P_0$ and the ratio in Eq.~\eqref{eq:expansion} are analytic in $\gamma^2$ in a neighbourhood of the XXZ line.

The Jacobi expansion at $m = (k')^2 = 1$ (Appendix~\ref{app:jacobi}) gives
\begin{align}\label{eq:Fexp}
F(\eta, k) &= \cosh\eta + k^2 F_2(\eta) + O(k^4), \nonumber\\
F_2(\eta) &= \frac{\sinh\eta(3\sinh\eta\cosh\eta - \eta)}{4}.
\end{align}
Writing $\eta(k) = \eta_0 + \eta_2 k^2 + O(k^4)$ and enforcing $F(\eta(k), k) = \cosh\eta_0$ at $O(k^2)$:
\begin{equation}\label{eq:eta2}
\eta_2 = -\frac{F_2(\eta_0)}{\sinh\eta_0} = \frac{\eta_0 - 3\sinh\eta_0\cosh\eta_0}{4}.
\end{equation}

The variable in~\eqref{eq:JMN} is $\eta_x \equiv \pi\lambda/[2K(k)]$ with $\lambda = \eta$ (Appendix~\ref{app:JMN-aux}), so it moves with $k$ through $\eta(k)$ and through $K(k)$:
\begin{align}\label{eq:eta_x_expansion}
\eta_x(k) &= \eta(k)(1 - k^2/4 + O(k^4)) \nonumber\\
&= \eta_0 + \left(\eta_2 - \frac{\eta_0}{4}\right)k^2 + O(k^4).
\end{align}
Substituting $\eta_2$ from~\eqref{eq:eta2}, the $\eta_0$-linear pieces cancel:
\begin{align}\label{eq:dexx}
\eta_2 - \frac{\eta_0}{4} &= -\frac{3\sinh\eta_0\cosh\eta_0}{4}, \nonumber\\
\partial_{\gamma^2}\eta_x\Big|_0 &= -\frac{3\cosh\eta_0}{4\sinh^3\eta_0},
\end{align}
where the conversion from $k^2$ to $\gamma^2$ uses Eq.~\eqref{eq:k2_conversion}. At fixed $x$ only the nome moves, and $P_0$ changes through $f(q)$ alone,
\begin{equation}\label{eq:naive}
\partial_{\gamma^2}\log f(q)\Big|_0 = \frac{1}{4(\Dz^2-1)^2} > 0,
\end{equation}
which is the first term of~\eqref{eq:CM}. The shift of $\eta_x$ adds the $g(x)$ contribution $\partial_\eta\log g(e^{-\eta})\Big|_{\eta_0}\cdot\partial_{\gamma^2}\eta_x\Big|_0$, which outweighs the nome term, so the full coefficient is negative; the negativity argument below makes this quantitative and identifies which part of the shift is responsible. The same implicit $\gamma$-dependence of the elliptic modulus and of $K(k)$ has to be tracked whenever an eight-vertex result is re-expressed in Hamiltonian variables, as it is in the entanglement-entropy calculations of Refs.~\cite{Ercolessi2010, Ercolessi2011}.

Because the ratio in Eq.~\eqref{eq:expansion} equals unity at $\gamma=0$, its $\gamma^2$ coefficient is $C_M=\partial_{\gamma^2}\log P_0|_0$. From Eqs.~\eqref{eq:logf}, \eqref{eq:nome_derivation}, and~\eqref{eq:k2_conversion},
\begin{equation}\label{eq:dlogf}
\partial_{\gamma^2}\log f(q)\Big|_0 = \frac{1}{4\sinh^4\eta_0}.
\end{equation}
From $\log g(e^{-\eta}) = 2\sum_n\log\tanh(n\eta)$ and the identity $\coth x - \tanh x = 2/\sinh(2x)$,
\begin{align}\label{eq:dgg}
\frac{\partial \log g(e^{-\eta})}{\partial\eta} &= 2\sum_n n[\coth(n\eta) - \tanh(n\eta)] \nonumber\\
&= 4\sum_{n=1}^\infty\frac{n}{\sinh(2n\eta)}.
\end{align}
Combining via the chain rule, the $\eta_x$-shift contribution is
\begin{align}\label{eq:etax_contribution}
\partial_\eta\log g(e^{-\eta})\Big|_{\eta_0}\partial_{\gamma^2}\eta_x\Big|_0 \qquad&\nonumber\\
= -\frac{3\cosh\eta_0}{\sinh^3\eta_0}&\sum_{n=1}^\infty\frac{n}{\sinh(2n\eta_0)},
\end{align}
and adding~\eqref{eq:dlogf},
\begin{align}\label{eq:CM_eta}
C_M = \frac{1}{4\sinh^4\eta_0} - \frac{3\cosh\eta_0}{\sinh^3\eta_0}\sum_{n=1}^\infty\frac{n}{\sinh(2n\eta_0)},
\end{align}
which becomes~\eqref{eq:CM} on substituting $\sinh^2\eta_0 = \Dz^2 - 1$, $\cosh\eta_0 = \Dz$.

It remains to establish the negativity bound. With $\eta = \arccosh\Dz > 0$, all terms in $\sum_n n/\sinh(2n\eta)$ are positive, so keeping only $n=1$,
\begin{equation}
\sum_{n=1}^\infty\frac{n}{\sinh(2n\eta)} > \frac{1}{\sinh(2\eta)} = \frac{1}{2\sinh\eta\cosh\eta}.
\end{equation}
Substituting into~\eqref{eq:CM_eta},
\begin{equation}
C_M < \frac{1}{4\sinh^4\eta} - \frac{3\cosh\eta}{\sinh^3\eta}\cdot\frac{1}{2\sinh\eta\cosh\eta}
= -\frac{5}{4\sinh^4\eta},
\end{equation}
and $\sinh^4\eta = (\Dz^2-1)^2$, which gives $C_M(\Dz) < -5/[4(\Dz^2-1)^2] < 0$ for all $\Dz > 1$.

The same estimate separates the two parts of the shift of $\eta_x$ in Eq.~\eqref{eq:eta_x_expansion}. Write $\Sigma=\sum_n n/\sinh(2n\eta_0)$. The part $\eta_2$, the shift of the crossing parameter at fixed $\Dz$, contributes $4\eta_2\Sigma/\sinh^4\eta_0$ to $C_M$, and the $K(k)$ rescaling contributes $-\eta_0\Sigma/\sinh^4\eta_0$. The nome term and the first part together give $C_M+\eta_0\Sigma/\sinh^4\eta_0<-5/(4\sinh^4\eta_0)+\eta_0/(2\sinh^5\eta_0\cosh\eta_0)$, which is negative because $\sinh2\eta_0>2\eta_0$, so the shift of the crossing parameter outweighs the nome term by itself. The $K(k)$ rescaling cancels the $\eta_0$-linear part of $\eta_2$ and changes the magnitude of $C_M$: near the isotropic point it supplies $-\pi^2/(16\eta_0^5)$ of the leading term $-3\pi^2/(16\eta_0^5)$ in Eq.~\eqref{eq:critical_eta}.

\section{Asymptotic behaviour}\label{sec:asymp}

In the large-$\Dz$ limit, with $x = e^{-\eta} = \Dz - \sqrt{\Dz^2-1} \to 0$, the hyperbolic functions take the forms
\begin{equation}\label{eq:hyperbolic_in_x}
\sinh\eta = \frac{1-x^2}{2x}, \quad \cosh\eta = \frac{1+x^2}{2x}.
\end{equation}
The Lambert sum admits the expansion (Appendix~\ref{app:lambert})
\begin{align}\label{eq:lambert_smallx}
\sum_{n=1}^\infty\frac{n}{\sinh(2n\eta)} &= 2\sum_{n=1}^\infty\frac{n x^{2n}}{1 - x^{4n}} \nonumber\\
&\hspace{-6em}= 2x^2 + 4x^4 + 8x^6 + 8x^8 + 12x^{10} + O(x^{12}),
\end{align}
where the coefficient of $x^{2N}$ is $2\sum_{d\mid N,\, N/d\text{ odd}} d$. Combining this with
\begin{align}\label{eq:two_prefactors}
\frac{1}{4\sinh^4\eta} &= \frac{4 x^4}{(1-x^2)^4}, &
\frac{3\cosh\eta}{\sinh^3\eta} &= \frac{12 x^2(1+x^2)}{(1-x^2)^3},
\end{align}
one obtains
\begin{align}\label{eq:largeDz_x}
C_M = -20 x^4 - 128 x^6 - 464 x^8 - 1216 x^{10} \nonumber\\
- 2620 x^{12} + O(x^{14}).
\end{align}
From the relation $x + 1/x = 2\Dz$,
\begin{equation}\label{eq:x_in_Delta}
x = \frac{1}{2\Dz} + \frac{1}{8\Dz^3} + \frac{1}{16\Dz^5}
+O(\Dz^{-7}),
\end{equation}
which on substitution into~\eqref{eq:largeDz_x} yields
\begin{align}\label{eq:largeDz_Delta}
C_M(\Dz) = -\frac{5}{4\Dz^4} - \frac{13}{4\Dz^6} - \frac{189}{32\Dz^8} + O(\Dz^{-10}),
\end{align}
as $\Dz \to \infty$. The negativity bound is therefore sharp in this limit: the part of the Lambert sum with $n\geq2$ is $O(x^4)$ and its prefactor in Eq.~\eqref{eq:two_prefactors} is $O(x^2)$, so it affects $C_M$ only from $O(x^6)$ onward, and the $n=1$ term fixes the coefficient $-20x^4=-5/(4\Dz^4)+O(\Dz^{-6})$.

In the opposite limit $\Dz \to 1^+$, the Eisenstein representation~\eqref{eq:eis_prev} combined with the modular transformation $E_2(-1/\tau) = \tau^2 E_2(\tau) - 6i\tau/\pi$ and the limit $E_2(i\infty) = 1$ gives (Appendix~\ref{app:lambert})
\begin{align}\label{eq:Lambert_asymp}
\sum_{n=1}^\infty\frac{n}{\sinh(2n\eta)} = \frac{\pi^2}{16\eta^2} - \frac{1}{4\eta} + \frac{\pi^2}{2\eta^2}e^{-\pi^2/\eta} \nonumber\\
+ O(\eta^{-2}e^{-2\pi^2/\eta}), \qquad \eta \to 0^+.
\end{align}
The remainder beyond the two power-law terms of the Lambert sum is exponentially small. All algebraic terms of the near-critical expansion of $C_M$ are therefore generated by the Laurent expansions of the two hyperbolic prefactors in~\eqref{eq:CM_eta}. These series converge for $0<|\eta|<\pi$, the nearest nonzero poles being at $\eta=\pm i\pi$. Explicitly,
\begin{align}
\frac{1}{4\sinh^4\eta} &= \frac{1}{4\eta^4}-\frac{1}{6\eta^2}
+\frac{11}{180}-\frac{31}{1890}\eta^2+O(\eta^4),\nonumber\\
\frac{3\cosh\eta}{\sinh^3\eta} &= \frac{3}{\eta^3}-\frac{\eta}{5}
+\frac{4\eta^3}{63}+O(\eta^5).
\end{align}
Multiplication by Eq.~\eqref{eq:Lambert_asymp} then gives
\begin{align}\label{eq:critical_eta}
C_M = -\frac{3\pi^2}{16\eta_0^5} + \frac{1}{\eta_0^4} - \frac{1}{6\eta_0^2} + \frac{\pi^2}{80\eta_0} + \frac{1}{90} + O(\eta_0),
\end{align}
as $\eta_0 \to 0^+$,
where the omitted modular contribution is $O(\eta_0^{-5}e^{-\pi^2/\eta_0})$. The $\eta_0^{-3}$ coefficient vanishes because neither prefactor contains the power needed to produce it. Continuing the multiplication gives the next three terms $-\frac{\pi^2}{252}\eta_0-\frac{1}{1890}\eta_0^2+\frac{\pi^2}{1200}\eta_0^3$. In terms of $\delta = \Dz - 1$, using $\eta_0 = \sqrt{2\delta}(1 - \delta/12 + \cdots)$,
\begin{align}\label{eq:critical_delta}
C_M(\Dz) = -\frac{3\pi^2}{64\sqrt{2}}\delta^{-5/2} + \frac{1}{4}\delta^{-2} \nonumber\\
- \frac{5\pi^2}{256\sqrt{2}}\delta^{-3/2} + O(\delta^{-1/2}),
\end{align}
as $\delta \to 0^+$, with no $\delta^{-1}$ term. The leading behaviour is
\begin{equation}\label{eq:critical}
C_M(\Dz) \sim -\frac{3\pi^2}{64\sqrt{2}}(\Dz-1)^{-5/2}, \qquad \Dz \to 1^+,
\end{equation}
with prefactor $-3\pi^2/(64\sqrt 2) \approx -0.327$.

For the joint limit, define
\begin{equation}
\ell(\eta)=\log g(e^{-\eta}).
\end{equation}
Since $\ell'(\eta)$ is four times the Lambert sum, Eq.~\eqref{eq:dgg}, integration of Eq.~\eqref{eq:Lambert_asymp} gives
\begin{equation}\label{eq:logg_asymptotic}
\ell(\eta)=-\frac{\pi^2}{4\eta}-\log\eta+\log(2\pi)+2e^{-\pi^2/\eta}
+O(e^{-2\pi^2/\eta}),
\end{equation}
where the constant follows, for example, from
$g(e^{-\eta})=\eta_{\rm D}(i\eta/\pi)^4/\eta_{\rm D}(2i\eta/\pi)^2$ and the modular transformation of the Dedekind eta function $\eta_{\rm D}$. Near $(\eta,k)=(0,0)$,
\begin{align}
F(\eta,k)&=1+\frac{1+k^2}{2}\eta^2+O(\eta^4),\nonumber\\
\cosh\eta_0&=1+\frac{\eta_0^2}{2}+O(\eta_0^4).
\end{align}
The analytic implicit-function theorem applied to $\eta^2$ therefore makes $\eta^2/\eta_0^2$, and hence its positive square root $\eta/\eta_0$, analytic in $\eta_0^2$ and $k^2$. Choosing this branch and using Eq.~\eqref{eq:dexx} gives the uniform expansions
\begin{align}\label{eq:uniform_dictionary}
\gamma &= k\eta_0^2[1+O(\eta_0^2+k^2)],\nonumber\\
\eta_x &=\eta_0\left[1-\frac{3}{4}k^2+O(\eta_0^2k^2+k^4)\right].
\end{align}
The error term in the second line follows from
$\eta_x-\eta_0=-3k^2\sinh\eta_0\cosh\eta_0/4+O(k^4)$ and the analyticity of $\eta_x/\eta_0$ at the origin. Equations~\eqref{eq:logf}, \eqref{eq:nome_derivation}, \eqref{eq:logg_asymptotic}, and \eqref{eq:uniform_dictionary} yield
\begin{align}\label{eq:uniform_log_ratio}
\log\frac{\Mz(\gamma,\cosh\eta_0)}{\Mz(0,\cosh\eta_0)}
&=-\frac{3\pi^2}{16}\frac{k^2}{\eta_0}
+O\left(k^2+\frac{k^4}{\eta_0}\right).
\end{align}
The estimate is uniform as $(\eta_0,k)\to(0,0)$ in a fixed sufficiently small neighbourhood. Now set $\gamma=\omega\eta_0^{5/2}$, with $\omega$ in a fixed compact set. The first line of Eq.~\eqref{eq:uniform_dictionary} gives
\begin{equation}
k=\omega\eta_0^{1/2}[1+O(\eta_0(1+\omega^2))],
\end{equation}
and hence
\begin{align}\label{eq:double_scaling_proof}
\log\frac{\Mz(\omega\eta_0^{5/2},\cosh\eta_0)}{\Mz(0,\cosh\eta_0)}
=-\frac{3\pi^2}{16}\omega^2
+O(\eta_0\omega^2(1+\omega^2)).
\end{align}
This proves Eq.~\eqref{eq:double_scaling_main}.

The exponent $5/2$ has two sources. The coupling relation~\eqref{eq:uniform_dictionary} gives $\gamma\simeq k\eta_0^2$. The remaining half power comes from the essential singularity of Baxter's formula: $\ell'(\eta)\simeq\pi^2/(4\eta^2)$ turns the shift $\eta_x-\eta_0\simeq-3k^2\eta_0/4$ into $-3\pi^2k^2/(16\eta_0)$, which is of order unity when $k$ is of order $\eta_0^{1/2}$. The scaling variable is therefore a power of $\eta_0$, while $\Mz(0,\Dz)$, Eq.~\eqref{eq:logg_asymptotic}, and the inverse correlation length~\cite{BaxterBook} vanish faster than any power. Written through the correlation length $\xi$, the crossover scale is $\gamma\propto(\ln\xi)^{-5/2}$.

If $\omega\to0$ as $\eta_0\to0$, exponentiation gives
\begin{equation}\label{eq:controlled_quadratic}
\frac{\Mz(\gamma,\cosh\eta_0)}{\Mz(0,\cosh\eta_0)}
=1-\frac{3\pi^2}{16}\omega^2
+O(\omega^4+\eta_0\omega^2(1+\omega^2)),
\end{equation}
so the quadratic truncation is controlled by the sufficient asymptotic condition $|\gamma|/\eta_0^{5/2}\to0$. Requiring the quadratic term to be small compared with unity, with $\eta_0^2=2(\Dz-1)[1+O(\Dz-1)]$, gives
\begin{equation}\label{eq:window}
|\gamma| \ll \left(\frac{64\sqrt{2}}{3\pi^2}\right)^{1/2}(\Dz-1)^{5/4}
[1+O(\Dz-1)].
\end{equation}
The scaling regime lies inside the N\'eel phase: the boundary $|\gamma|=\Dz-1=\eta_0^2/2+O(\eta_0^4)$ of Sec.~\ref{sec:full} corresponds to $|\omega|=\eta_0^{-1/2}/2+O(\eta_0^{3/2})$, which leaves any fixed compact set of $\omega$ once $\eta_0$ is small enough. At moderate $\eta_0$ the two scales are not far apart; for $\omega=1$ the boundary is reached near $\eta_0=1/4$. For fixed nonzero $\omega$, the powers generated by expanding the exponential in Eq.~\eqref{eq:double_scaling_main} are of the same scaling order and are summed by that equation. A fixed-$\gamma$ approach to $\Dz=1$ lies outside this neighbourhood, so a continuum comparison can use either the iterated coefficient limit or the double-scaling function~\eqref{eq:double_scaling_main}.

\section{Full XY-anisotropy dependence and the phase boundary}\label{sec:full}

The coupling relations~\eqref{eq:algebraic_dictionary} and their inversion~\eqref{eq:k_exact} hold at any $\gamma$ in the antiferromagnetic phase. Since $\Mz$ is even in $\gamma$ we take $\gamma\geq0$. For $0\leq\gamma<\Dz-1$ the branch~\eqref{eq:k_exact} gives $0\leq k<1$, and Eq.~\eqref{eq:algebraic_dictionary} gives $\scn^2(\lambda;k')=\gamma/k$. With $\scn=\tan\operatorname{am}$, the crossing parameter is an incomplete elliptic integral of the first kind,
\begin{equation}\label{eq:lambda_full}
\lambda(\gamma,\Dz)=F\left(\arctan\sqrt{\gamma/k};k'\right), \qquad 0<\lambda<K',
\end{equation}
and the staggered magnetization is
\begin{equation}\label{eq:Mz_full}
\Mz(\gamma,\Dz)=f(q)g(x), \qquad q=e^{-\pi K'/K}, \quad x=e^{-\pi\lambda/(2K)},
\end{equation}
with $f$ and $g$ from Eq.~\eqref{eq:JMN}, $K=K(k)$, and $K'=K(k')$. As $\gamma\to0$, $\gamma/k\to\sinh^2\eta_0$ and $F(\arctan\sinh\eta_0;1)=\eta_0$, which returns the XXZ limit of Appendix~\ref{app:JMN-aux-C}. Equations~\eqref{eq:k_exact}, \eqref{eq:lambda_full}, and~\eqref{eq:Mz_full} give $\Mz$ in the Hamiltonian variables with no implicit step; the coefficient $C_M$ of Sec.~\ref{sec:derivation} is their $\gamma^2$ Taylor coefficient.

\begin{figure}[t]
  \centering
\includegraphics[width=\columnwidth]{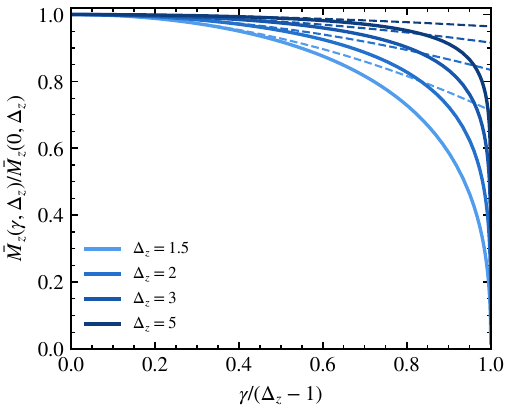}
  \caption{The ratio $\Mz(\gamma,\Dz)/\Mz(0,\Dz)$ from Eqs.~\eqref{eq:lambda_full} and~\eqref{eq:Mz_full} (solid) and its quadratic truncation $1+C_M(\Dz)\gamma^2$ (dashed), against $\gamma/\gamma_c$ with $\gamma_c=\Dz-1$, for $\Dz=1.5$, $2$, $3$, and $5$ (light to dark).}
  \label{fig:Mz_full}
\end{figure}

The square root in Eq.~\eqref{eq:k_exact} vanishes when $\sinh^2\eta_0-\gamma^2=2\gamma$, that is at
\begin{equation}\label{eq:gamma_c}
\gamma_c=\Dz-1,
\end{equation}
where $k=1$ and both $q$ and $x$ tend to unity. In Eq.~\eqref{eq:H}, $\Dz$ is the coupling of largest magnitude for $|\gamma|<\Dz-1$; at $\gamma=\gamma_c$ the $x$ and $z$ couplings are equal, and the chain is a gapless XXZ chain with symmetry axis $y$ and anisotropy
\begin{equation}\label{eq:mu_def}
\cos\mu=\frac{1-\gamma_c}{1+\gamma_c}=\frac{2-\Dz}{\Dz}, \qquad 0<\mu<\pi.
\end{equation}
For $\gamma>\Dz-1$ the order is along $x$~\cite{BaxterBook, denNijs1981}. Figure~\ref{fig:Mz_full} shows $\Mz(\gamma,\Dz)/\Mz(0,\Dz)$ across the phase. The quadratic truncation~\eqref{eq:expansion} follows the curve over roughly the inner half of the phase and lies above it further out. Near $\gamma_c$ the order parameter vanishes as a power law, which no finite truncation in $\gamma^2$ reproduces.

The behaviour at the boundary follows from a modular transformation of both factors in Eq.~\eqref{eq:JMN}. Writing the $q$-Pochhammer symbols through the Dedekind eta function, $f(q)=\eta_{\rm D}(2\tau)^2/\eta_{\rm D}(\tau)^4$ with $\tau=iK'/(2K)$, and $g(x)=\eta_{\rm D}(i\eta_x/\pi)^4/\eta_{\rm D}(2i\eta_x/\pi)^2$ as in Sec.~\ref{sec:asymp}. The transformation $\eta_{\rm D}(-1/\tau)=\sqrt{-i\tau}\eta_{\rm D}(\tau)$, applied to both, gives
\begin{equation}\label{eq:P0_modular}
P_0=\frac{K'}{\lambda}\exp\left[-\frac{\pi K}{2}\left(\frac{1}{\lambda}-\frac{1}{K'}\right)\right]\left[1+O(e^{-2\pi K/K'})\right],
\end{equation}
where the remainder comes from the $f$ factor; the one from $g$, $O(e^{-2\pi K/\lambda})$, is smaller since $\lambda<K'$. As $\gamma\to\gamma_c^-$, $k'\to0$, $K=\ln(4/k')+o(1)$, $K'\to\pi/2$, and $\lambda\to\lambda_c=\arctan\sqrt{\Dz-1}$, with $2\lambda_c=\mu$ by Eq.~\eqref{eq:mu_def}. Equation~\eqref{eq:k_equation} gives $k'^4=8\Dz(\gamma_c-\gamma)/(\Dz-1)[1+o(1)]$. The corrections to $K'$ and $\lambda$ are $O(k'^2)$ and enter the exponent of Eq.~\eqref{eq:P0_modular} multiplied by $K=O(\ln k')$, so they vanish in the limit, and
\begin{equation}\label{eq:Mz_boundary}
\Mz(\gamma,\Dz)\simeq\frac{\pi}{\mu}\left[\frac{\Dz(\gamma_c-\gamma)}{32(\Dz-1)}\right]^{\beta}, \qquad \beta=\frac{\pi-\mu}{4\mu},
\end{equation}
as $\gamma\to\gamma_c^-$, with a relative correction $O\left((\gamma_c-\gamma)^{1/2}\ln(\gamma_c-\gamma)\right)$.

The exponent depends on $\Dz$ only through the anisotropy of the critical chain on the boundary, and it agrees with the scaling estimate for an XXZ critical line perturbed by in-plane anisotropy~\cite{LutherPeschel1975, denNijs1981}. With Luttinger parameter $K_{\rm L}=\pi/[2(\pi-\mu)]$, normalized to $K_{\rm L}=1$ at the XX point and $K_{\rm L}=1/2$ at the isotropic point, the staggered component of $\sz$ has scaling dimension $1/(4K_{\rm L})$ and the perturbation $J_z-J_x\propto\gamma_c-\gamma$ has dimension $1/K_{\rm L}$~\cite{Giamarchi2004}. The order parameter then scales as $(\gamma_c-\gamma)^{\beta}$ with $\beta=1/[4K_{\rm L}(2-1/K_{\rm L})]$, which is the exponent of Eq.~\eqref{eq:Mz_boundary}. At $\Dz=2$ the boundary chain is the XX chain and $\beta=1/4$. As $\Dz\to1^+$, $\mu\to0$ and $\beta$ diverges, in line with the essential singularity of the XXZ order parameter at the isotropic point; as $\Dz\to\infty$, $\beta\to0$.

\section{Numerical comparison}\label{sec:numerics}

The infinite-DMRG calculations use a 2-site unit cell and no enforced $U(1)$ symmetry, which $\gamma\neq0$ breaks. The Hamiltonian is implemented in Pauli normalization, cf.~the remark below Eq.~\eqref{eq:H}, and the N\'eel-up symmetry-broken sector is selected by product-state initialization $| \uparrow\downarrow\rangle$. Convergence is checked by requiring both the variational energy and the bipartite entanglement entropy to be stable to at least ten significant digits across a minimum of ten consecutive sweeps, with a singular-value cutoff of $10^{-14}$; the resulting uncertainty on $\Mz$ itself is at the $10^{-8}$ level or below, well below the $\gamma^2$ signal extracted here\footnote{The longitudinal two-point function was computed by two independent methods --- direct transfer-matrix contraction and sequential one-site expectation evaluation along a chain --- with the two implementations agreeing at all $\Dz$. Under-converged iDMRG can produce non-canonical iMPS tensors whose two-point correlators diverge with distance even when one-point functions appear converged; the boundedness $|\langle\sz_0\sz_n\rangle|\leq 1$ is a simple diagnostic.}. The bond dimensions are listed in Table~\ref{tab:validation}; a VUMPS optimizer~\cite{ZaunerStauber2018} can be used in the same way. The order parameter is extracted as
\begin{equation}\label{eq:Mz_extr}
\Mz = \frac{1}{2}|\langle\sz_A\rangle - \langle\sz_B\rangle|,
\end{equation}
with $A, B$ the two sites of the unit cell (the symmetry $\langle\sz_A\rangle = -\langle\sz_B\rangle$ holds in all runs). For each $\Dz$ we run iDMRG at $\gamma = 0$ and three to four small positive $\gamma$, and compute
\begin{equation}\label{eq:cm_extraction}
C_M^{\rm iDMRG}(\gamma) = \frac{1}{\gamma^2}\log\frac{\Mz(\gamma)}{\Mz(0)},
\end{equation}
linearly extrapolated to $\gamma^2 \to 0$. The upper end of each $\gamma$-range is chosen so that the $O(\gamma^6)$ contribution to $\log[\Mz(\gamma)/\Mz(0)]$ remains below the iDMRG resolution, as verified by the linearity of $C_M^{\rm iDMRG}(\gamma)$ in $\gamma^2$. The extrapolated values are collected in Table~\ref{tab:validation}. An uncertainty $\delta\Mz$ in each magnetization enters $C_M^{\rm iDMRG}(\gamma)$ as roughly $2\delta\Mz/(\Mz\gamma^2)$, so the smallest $\gamma$ of each set limits the resolution; at $\Dz=1.5$, where the $\gamma$-range is narrowest, this is of the same order as the difference between the two values of $C_M$ in the table. Two further checks, independent of the iDMRG implementation and of each other, corroborate the closed form at the percent level attainable within their schemes: an iTEBD calculation at $\Dz=3$, in which the order parameter is measured by transfer-matrix contraction after imaginary-time evolution, and exact diagonalization on 14-site rings at $\Dz=6$ and $8$, which extracts $\Mz$ from the long-distance staggered correlator with no symmetry-breaking field.

\begin{table}[b]
\centering
\footnotesize
\setlength{\tabcolsep}{3.5pt}
\begin{tabular}{cccccc}
\toprule
$\Dz$ & $\Mz(0)$ & $C_M$, Eq.~\eqref{eq:CM} & $C_M^{\rm iDMRG}$ & $\chi$ & $\gamma_{\max}$ \\
\midrule
$1.5$ & $0.50284$ & $-1.1485$ & $-1.1493$ & $80$ & $0.012$ \\
$2.0$ & $0.73352$ & $-0.16553$ & $-0.16557$ & $80$ & $0.06$ \\
$3.0$ & $0.88572$ & $-0.020973$ & $-0.020977$ & $40$ & $0.07$ \\
$5.0$ & $0.95960$ & $-0.002224$ & $-0.002224$ & $20$ & $0.12$ \\
\bottomrule
\end{tabular}
\caption{iDMRG comparison for $C_M(\Dz)$. For each $\Dz$, runs at $\gamma \in [0, \gamma_{\max}]$ enter the extrapolation~\eqref{eq:cm_extraction}. The unperturbed magnetization $\Mz(0)$ reproduces the Baxter XXZ formula at every $\Dz$.}
\label{tab:validation}
\end{table}

The data in Table~\ref{tab:validation} cover four representative anisotropies in the moderately and strongly gapped regime $\Dz \geq 1.5$, where $\eta_0 \gtrsim 1$ and the bond dimensions reported suffice to converge $\Mz$ to the precision shown. A quantitative test of the expansion~\eqref{eq:critical_delta} requires data at $\Dz - 1 \ll 1$, where the growing correlation length forces substantially larger bond dimensions and the window~\eqref{eq:window} narrows.

\section{Discussion}\label{sec:discussion}

The closed form~\eqref{eq:CM} follows from the spontaneous polarization of Refs.~\cite{BaxterKelland1974, JMN1993} combined with the parametrization of Appendix~\ref{app:JMN-aux}, expanded at fixed longitudinal anisotropy. The polarization itself has long been known in elliptic form, and is reproduced in vertex-operator language by the free-field construction of Refs.~\cite{LashkevichPugai1998, Lashkevich2002}, within the massive-regime framework of Refs.~\cite{DFJMN1993, JimboMiwaBook}. The step added here is the algebraic inversion of the coupling relations, which turns the polarization into a function of the Hamiltonian couplings over the whole N\'eel phase, and its differentiation at fixed couplings. Both shifts of the crossing-parameter variable enter that differentiation. The implicit shift of the crossing parameter dominates the coefficient and fixes its sign; the rescaling by the complete elliptic integral cancels the $\eta_0$-linear part of that shift and contributes a third of the near-critical divergence.

One place to look for an independent cross-check is the supersymmetric, or combinatorial, XYZ line~\cite{RazumovStroganov2010, HagendorfFendley2012, HagendorfRosengren2024, BazhanovMangazeev2007, BazhanovMangazeev2010, Zinn2013}, on which finite-$N$ correlation functions are known in terms of Painlev\'e-VI $\tau$-functions. Its defining condition is not invariant under the sublattice rotations that relate equivalent XYZ chains. For the sign pattern that reaches the present phase, the line meets each fixed-$\Dz$ section at the single point $\gamma^*=\sqrt{\Dz^2+1}-\Dz$, which lies inside the N\'eel phase for $\Dz>4/3$ (Appendix~\ref{app:susy}). There the nome of the polarization formula is the sixth power of the auxiliary variable, so the polarization reduces to a function of one variable, and simple expressions for the order parameters on this line were conjectured in Ref.~\cite{FendleyHagendorf2010}.

In the critical direction, the divergence of $C_M$ with its subleading terms, Eq.~\eqref{eq:critical_delta}, and the scaling function~\eqref{eq:double_scaling_main} are the lattice data for a continuum sine-Gordon matching of the staggered moment near the gapless Heisenberg point. The relevant continuum object is the response of the sine-Gordon order-parameter expectation value to the dual cosine deformation that represents the XY anisotropy~\cite{LukyanovZamolodchikov1997}. The undeformed expectation value is known explicitly~\cite{LukyanovZamolodchikov1997}. Finally, deriving its response would connect the continuum result to both the fixed-$\Dz$ coefficients and the joint scaling limit obtained here.

The quartic coefficient of Eq.~\eqref{eq:expansion} is accessible by carrying the elliptic expansion one order further. A high-bond-dimension iDMRG or VUMPS~\cite{ZaunerStauber2018} study close to the critical point could test the near-critical expansion~\eqref{eq:critical_delta} coefficient by coefficient within the window~\eqref{eq:window}, its subleading coefficients being parameter-free targets. The same calculations along lines of fixed longitudinal anisotropy towards the phase boundary would test the boundary law~\eqref{eq:Mz_boundary}, where the correlation length again grows. A derivation of the first-order term in the nome within the free-field screening construction~\cite{LashkevichPugai1998, Lashkevich2002} would reproduce the rational part of Eq.~\eqref{eq:CM} algebraically. The $Z_n$-Baxter generalizations of the spontaneous polarization~\cite{Quano1993} and the higher-rank $U_q(\widehat{sl}(n))$ extensions~\cite{Koyama1994} provide a wider family in which an analogous Hamiltonian-variable expansion may be tractable.

\section*{Data availability}

The data that support the findings of this article are available from the author upon reasonable request.

\appendix

\section{Jacobi elliptic derivatives at \texorpdfstring{$m=1$}{m=1}}\label{app:jacobi}

We derive~\eqref{eq:Fexp}. Set $m = (k')^2 = 1 - k^2$, so $\partial_{k^2} = -\partial_m$. From $(\sn')^2 = (1-\sn^2)(1-m\sn^2)$ at $m = 1$ (where $\sn(u\mid 1) = \tanh u$, $\cn(u\mid 1) = \dn(u\mid 1) = \operatorname{sech} u$), with $\dot s(u) \equiv \partial_m\sn(u\mid m)|_{m=1}$:
\begin{equation}
(\dot s)' + 2\dot s\tanh u = -\frac{1}{2}\tanh^2 u, \qquad \dot s(0) = 0.
\end{equation}
With integrating factor $\cosh^2 u$,
\begin{equation}\label{eq:sdot}
\dot s(u) = \frac{u - \sinh u\cosh u}{4\cosh^2 u}.
\end{equation}
From $\sn^2+\cn^2 = 1$ and $\dn^2 = 1 - m\sn^2$,
\begin{align}\label{eq:cdot_ddot}
\dot c(u) &= \frac{\sinh u(\sinh u\cosh u - u)}{4\cosh^2 u}, \nonumber\\
\dot d(u) &= -\frac{\sinh u(\sinh u\cosh u + u)}{4\cosh^2 u}.
\end{align}
For $F(\eta, k) = \dn/\cn^2$ at fixed $\eta$, $\partial_{k^2} F\Big|_0 = -\partial_m F\Big|_{m=1} = \cosh^2\eta(2\dot c - \dot d)$, so
\begin{equation}\label{eq:F2_clean}
F_2(\eta) = \cosh^2\eta(2\dot c - \dot d)\Big|_{u=\eta} = \frac{\sinh\eta(3\sinh\eta\cosh\eta - \eta)}{4}.
\end{equation}
Standard tabulations of $\sn, \cn, \dn$ Taylor coefficients in $m$ at $m = 1$ are in~\cite{WhittakerWatson, Lawden1989}. The particular combination $F_2 = \cosh^2\eta(2\dot c - \dot d)$ that arises in the XYZ-chain expansion does not appear to be tabulated.

\section{Modular identity and asymptotics of the Lambert sum}\label{app:lambert}

To derive the Eisenstein representation, set $t = e^{-2\eta}$ (a dummy expansion variable, distinct from the eight-vertex nome $q$ of the main text) and use $1/\sinh(2n\eta) = 2t^n/(1-t^{2n})$:
\begin{equation}\label{eq:Lambert_q}
\sum_{n=1}^\infty\frac{n}{\sinh(2n\eta)} = 2\sum_{n=1}^\infty\frac{n t^n}{1-t^{2n}}.
\end{equation}
Using the partial-fraction and telescoping identities
\begin{align}\label{eq:two_identities}
\frac{1}{1-t^{2n}} &= \frac{1}{2}\left[\frac{1}{1-t^n} + \frac{1}{1+t^n}\right], \nonumber\\
\frac{t^n}{1+t^n} &= \frac{t^n}{1-t^n} - \frac{2t^{2n}}{1-t^{2n}},
\end{align}
one obtains
\begin{align}
\sum_n\frac{n t^n}{1-t^{2n}} &= \sum_n\frac{n t^n}{1-t^n} - \sum_n\frac{n t^{2n}}{1-t^{2n}} \nonumber\\
&= \frac{1-E_2(\tau)}{24} - \frac{1-E_2(2\tau)}{24} \nonumber\\
&= \frac{E_2(2\tau) - E_2(\tau)}{24},
\end{align}
where $E_2(\tau) = 1 - 24\sum_{n\geq 1} \sigma_1(n) t^n$ (with $t = e^{2\pi i\tau}$ and $\sigma_1(n)$ the divisor sum) is the quasi-modular weight-$2$ Eisenstein series, and $\tau = i\eta/\pi$. Therefore
\begin{equation}\label{eq:Eis_identity}
\sum_{n=1}^\infty\frac{n}{\sinh(2n\eta)} = \frac{E_2(2\tau) - E_2(\tau)}{12}.
\end{equation}
For the small-$\eta$ asymptotic, the modular transformation of $E_2$ is
\begin{equation}\label{eq:E2_modular}
E_2(-1/\tau) = \tau^2 E_2(\tau) - \frac{6 i \tau}{\pi}.
\end{equation}
Inverting it at $\tau = i\eta/\pi$ and at $2\tau = 2i\eta/\pi$, together with $E_2(i\pi/\eta) = 1 - 24e^{-2\pi^2/\eta} + O(e^{-4\pi^2/\eta})$ and $E_2(i\pi/(2\eta)) = 1 - 24e^{-\pi^2/\eta} + O(e^{-2\pi^2/\eta})$,
\begin{align}\label{eq:E2_pair}
E_2(i\eta/\pi) &= -\frac{\pi^2}{\eta^2} + \frac{6}{\eta} + \frac{24\pi^2}{\eta^2}e^{-2\pi^2/\eta} \nonumber\\
&\quad + O(\eta^{-2}e^{-4\pi^2/\eta}), \nonumber\\
E_2(2 i\eta/\pi) &= -\frac{\pi^2}{4\eta^2} + \frac{3}{\eta} + \frac{6\pi^2}{\eta^2}e^{-\pi^2/\eta} \nonumber\\
&\quad + O(\eta^{-2}e^{-2\pi^2/\eta}).
\end{align}
Inserting these into~\eqref{eq:Eis_identity},
\begin{align}\label{eq:Lambert_asymp_app}
\sum_{n=1}^\infty\frac{n}{\sinh(2n\eta)} = \frac{\pi^2}{16\eta^2} - \frac{1}{4\eta} + \frac{\pi^2}{2\eta^2}e^{-\pi^2/\eta} \nonumber\\
+ O(\eta^{-2}e^{-2\pi^2/\eta}), \qquad \eta\to 0^+.
\end{align}
Thus the algebraic part of the Lambert sum terminates after $-1/(4\eta)$. Multiplying the right-hand side by four and integrating, as Eq.~\eqref{eq:dgg} requires, gives Eq.~\eqref{eq:logg_asymptotic}; in particular, $\int 2\pi^2\eta^{-2}e^{-\pi^2/\eta}d\eta=2e^{-\pi^2/\eta}$. Multiplied by the convergent Laurent series of the hyperbolic prefactors, the result determines every algebraic coefficient of $C_M$; a finite truncation of the near-critical expansion carries the next algebraic term as its remainder, and only the summed series leaves an exponentially small one.

In the opposite, large-$\eta$ regime, with $x = e^{-\eta}$ and $t = x^2$, equation~\eqref{eq:Lambert_q} gives
\begin{equation}
\sum_n\frac{n}{\sinh(2n\eta)} = 2\sum_{n=1}^\infty\frac{n x^{2n}}{1 - x^{4n}} = \sum_{N\geq 1} a_N x^{2N},
\end{equation}
with $a_N = 2\sum_{d\mid N,\, N/d\text{ odd}} d$, giving $a_1 = 2$, $a_2 = 4$, $a_3 = 8$, $a_4 = 8$, $a_5 = 12$, $a_6 = 16$, and so on.

\section{Eight-vertex conventions and the Hamiltonian couplings}\label{app:JMN-aux}

This appendix fixes the conventions relating the eight-vertex parametrization of Ref.~\cite{JMN1993} to the Hamiltonian couplings of Eq.~\eqref{eq:H}, and derives Eq.~\eqref{eq:dict}. Two distinct parameters appear in the weight parametrization: a crossing (anisotropy) parameter, which we call $\lambda$ following~\cite{JMN1993}, and a spectral parameter $u$. The polarization depends on the former only.

\subsection{Weights, crossing parameter, and spectral parameter}\label{app:JMN-aux-A}

In the conventions of Ref.~\cite{JMN1993}, the Boltzmann weights of the eight-vertex model in the antiferroelectric regime are
\begin{align}\label{eq:jmn_weights}
a &= -i\rho\Theta(i\lambda)H(i\lambda-iu)\Theta(iu), \nonumber\\
b &= -i\rho\Theta(i\lambda)\Theta(i\lambda-iu)H(iu), \nonumber\\
c &= -i\rho H(i\lambda)\Theta(i\lambda-iu)\Theta(iu), \nonumber\\
d &= \phantom{-}i\rho H(i\lambda)H(i\lambda-iu)H(iu),
\end{align}
where $H$ and $\Theta$ are the Jacobi theta functions of modulus $k$ and quarter-periods $I = K(k)$, $I' = K'(k)$, $\lambda$ is the crossing parameter, and $u$ is the spectral parameter. The associated multiplicative variables are
\begin{equation}\label{eq:jmn_variables}
q = e^{-\pi I'/I}, \qquad x = e^{-\pi\lambda/(2I)}, \qquad \zeta = e^{\pi u/(2I)},
\end{equation}
with the antiferroelectric (regime-III) domain $0 < q < x < \zeta^{-1} < 1$, i.e.\ $0 < u < \lambda < 2I'$. Transfer matrices with common $(k, \lambda)$ and different $u$ commute, so $u$ parametrizes a commuting family while $(k,\lambda)$ label the model. The spontaneous staggered polarization proved in Ref.~\cite{JMN1993} is the function $P_0(x,q)$ of Eq.~\eqref{eq:JMN}: it depends on $q$ and on the exponentiated crossing parameter $x$ alone, and the spectral parameter drops out, as it must for a thermodynamic one-point function. The variable that we manipulate in Secs.~\ref{sec:derivation} and~\ref{sec:asymp} is therefore
\begin{equation}\label{eq:eta_x_def}
\eta_x \equiv \frac{\pi\lambda}{2K(k)}, \qquad x = e^{-\eta_x},
\end{equation}
as fixed by the parametrization~\eqref{eq:jmn_variables} in which the theorem of Ref.~\cite{JMN1993} is stated. We use $\eta=\lambda$ throughout; in the notation of Ref.~\cite{BaxterBook} this corresponds to $\eta_B=i\lambda/2$ and spectral variable $v=i\lambda/2-iu$.

\subsection{Hamiltonian point and coupling constants}\label{app:JMN-aux-B}

At $u = 0$ the weights~\eqref{eq:jmn_weights} degenerate: $H(0) = 0$ gives $b = d = 0$ and $a = c$, so the $R$-matrix
\begin{equation}\label{eq:Rmatrix}
R(u) = \begin{pmatrix} a & 0 & 0 & d\\ 0 & b & c & 0\\ 0 & c & b & 0\\ d & 0 & 0 & a\end{pmatrix}
\end{equation}
reduces at $u=0$ to $a(0)$ times the permutation operator $P$, and the transfer matrix becomes the lattice translation. The spin chain is obtained from the logarithmic derivative at this point,
\begin{equation}\label{eq:Hlimit}
H = -J_0\partial_u \log T(u)\Big|_{u=0} + \text{const}, \qquad J_0 > 0,
\end{equation}
the sign being fixed by the requirement that the eight-vertex equilibrium state correspond to the ground state of $H$~\cite{BaxterBook, Baxter1972}. The Pauli decomposition can be performed entry by entry. Since
\begin{align}
P\partial_uR(0)={}&\alpha_0\mathbf 1
+\frac{b'(0)+d'(0)}{2}\sx\otimes\sx\nonumber\\
&+\frac{b'(0)-d'(0)}{2}\sy\otimes\sy\nonumber\\
&+\frac{a'(0)-c'(0)}{2}\sz\otimes\sz,
\end{align}
where $\alpha_0=[a'(0)+c'(0)]/2$, only three combinations of derivatives are needed. Put $A_\lambda=\Theta(i\lambda)$, $B_\lambda=H(i\lambda)$, $\theta_0=\Theta(0)$, and $h_0=H'(0)$. Direct differentiation of Eq.~\eqref{eq:jmn_weights}, using $H(0)=\Theta'(0)=0$, gives
\begin{align}\label{eq:weight_derivatives}
b'(0)+d'(0)&=\rho h_0(A_\lambda^2-B_\lambda^2),\nonumber\\
b'(0)-d'(0)&=\rho h_0(A_\lambda^2+B_\lambda^2),\nonumber\\
a'(0)-c'(0)&=-\rho\theta_0
[A_\lambda H'(i\lambda)-B_\lambda\Theta'(i\lambda)].
\end{align}
Any $u$-dependence of the common normalization $\rho$ contributes only to $\alpha_0$ and hence only to the additive constant in Eq.~\eqref{eq:Hlimit}. The identity $H(w)=\sqrt{k}\sn(w;k)\Theta(w)$~\cite{WhittakerWatson, Lawden1989} gives
\begin{align}
A_\lambda^2\mp B_\lambda^2
&=A_\lambda^2[1\mp k\sn^2(i\lambda;k)],\nonumber\\
A_\lambda H'(i\lambda)-B_\lambda\Theta'(i\lambda)
&=\sqrt{k}A_\lambda^2\cn(i\lambda;k)\dn(i\lambda;k),
\end{align}
and $h_0=\sqrt{k}\theta_0$. The common factor in Eq.~\eqref{eq:weight_derivatives} cancels from the coupling ratios, leaving
\begin{align}\label{eq:Jratios}
J_x : J_y : J_z = \left[1 - k\sn^2(i\lambda)\right] : \left[1 + k\sn^2(i\lambda)\right] \nonumber\\
: -\cn(i\lambda)\dn(i\lambda),
\end{align}
before the overall sign of Eq.~\eqref{eq:Hlimit} is applied. The Jacobi imaginary transformation, $\sn(i\lambda;k) = i\scn(\lambda;k')$, $\cn(i\lambda;k) = 1/\cn(\lambda;k')$, $\dn(i\lambda;k) = \dn(\lambda;k')/\cn(\lambda;k')$, renders all three entries real:
\begin{equation}\label{eq:Jratios_real}
J_x : J_y : J_z = (1+\gamma) : (1-\gamma) : -\Dz,
\end{equation}
with
\begin{equation}\label{eq:dict_exact}
\gamma = k\scn^2(\lambda;k'), \qquad \Dz = \frac{\dn(\lambda;k')}{\cn^2(\lambda;k')},
\end{equation}
both valid at all $0<k<1$. The logarithmic derivative divides the combinations~\eqref{eq:weight_derivatives} by $a(0)=-i\rho\theta_0A_\lambda B_\lambda$; with $B_\lambda=i\sqrt{k}\scn(\lambda;k')A_\lambda$ the normalization $\rho$ drops out, and the three couplings share the factor $1/[2\scn(\lambda;k')]$, which is positive for $0<\lambda<I'$. Including the overall minus sign of Eq.~\eqref{eq:Hlimit} gives the chain $-J_0[(1+\gamma)\sx\sx + (1-\gamma)\sy\sy - \Dz\sz\sz]$; the sublattice rotation $U = \prod_{n\ \mathrm{odd}}\sz_n$, which reverses the sign of the $\sx\sx$ and $\sy\sy$ couplings and leaves $\sz_n$ (hence $\Mz$) invariant, then produces Eq.~\eqref{eq:H} with $J = J_0$ and $\eta \equiv \lambda$. Expanding~\eqref{eq:dict_exact} at small $k$ yields the first entry of Eq.~\eqref{eq:dict}, while Eq.~\eqref{eq:nome_derivation} supplies the second.

At fixed physical $\Dz$, Eq.~\eqref{eq:dict_exact} defines $\lambda(k)$ implicitly, and the variable entering the polarization is $\eta_x = \pi\lambda(k)/[2K(k)]$, whose numerator and denominator both move with $k$; Sec.~\ref{sec:derivation} carries out the expansion.

\subsection{Reduction to the Baxter XXZ formula at \texorpdfstring{$k=0$}{k=0}}\label{app:JMN-aux-C}

At $k = 0$ the parametrization collapses onto the XXZ chain: $k' = 1$, $K(0) = \pi/2$, $K'(0) = \infty$, and the nome $q$ tends to zero. The $q$-Pochhammer products in $f$ then all evaluate to unity, so $f(0) = 1$. The rescaling in $\eta_x = \pi\eta/(2K(0))$ becomes trivial, leaving $\eta_x = \eta$ and $x = e^{-\eta}$, the XXZ variable. Finally $(1 - e^{-2n\eta})/(1 + e^{-2n\eta}) = \tanh(n\eta)$ turns $g$ into $\prod_{n\geq 1}\tanh^2(n\eta)$. Combining, $P_0|_{k=0} = \prod_{n\geq 1}\tanh^2(n\eta_0) = \Mz^{\rm XXZ}(\eta_0)$, Baxter's classical formula~\cite{BaxterBook}, with $\eta_0 = \arccosh\Dz$.

\subsection{Supersymmetric line}\label{app:susy}

The supersymmetric line is defined by $J_x J_y + J_y J_z + J_z J_x = 0$. The condition is not invariant under the sublattice rotations that relate equivalent XYZ chains, so it has to be applied to each sign pattern of the couplings. Applied to the couplings of Eq.~\eqref{eq:H} as written, it gives $\gamma^2 = 1 + 2\Dz$, far from the XXZ line. After the rotation $\prod_{n\ \mathrm{odd}}\sy_n$, which reverses the signs of the $\sx\sx$ and $\sz\sz$ couplings, it gives $\gamma^2 + 2\gamma\Dz = 1$, whose root $\gamma^* = \sqrt{\Dz^2+1}-\Dz$ lies in the gapped N\'eel regime, $\gamma^*<\Dz-1$, for $\Dz > 4/3$ and approaches the XXZ line as $\gamma^*\simeq 1/(2\Dz)$ at large $\Dz$; the rotation $\prod_{n\ \mathrm{odd}}\sx_n$ gives its mirror image $-\gamma^*$. In the variables of Eq.~\eqref{eq:jmn_variables} this branch is the line $\lambda = I'/3$, that is $q = x^6$: there $\sn(2\lambda;k') = \operatorname{cd}(\lambda;k')$ and $\cn(2\lambda;k') = k\operatorname{sd}(\lambda;k')$, and the duplication formulas reduce Eq.~\eqref{eq:dict_exact} to $\gamma^2 + 2\gamma\Dz = 1$. Along it the polarization is the one-variable function $f(x^6)g(x)$.

\section{Strong-coupling check by lattice perturbation theory}\label{app:pert}

This appendix rederives the leading large-$\Dz$ behaviour of $C_M$,
\begin{equation}\label{eq:ptclaim}
C_M(\Dz) = -\frac{5}{4\Dz^4} + O(\Dz^{-6}),
\end{equation}
directly from the Hamiltonian~\eqref{eq:H} by Rayleigh--Schr\"odinger perturbation theory around the Ising limit. No eight-vertex input enters: the calculation is a self-contained check of Eq.~\eqref{eq:CM} in the regime where the negativity bound of Sec.~\ref{sec:derivation} becomes sharp, and it identifies the virtual processes responsible for the suppression of the staggered moment.

\subsection{Setup}\label{app:pt-setup}

Set $J = 1$ and split Eq.~\eqref{eq:H} as $H = H_0 + T + \gamma V$ with
\begin{align}\label{eq:pt_split}
H_0 &= \Dz\sum_n \sz_n\sz_{n+1}, \nonumber\\
T &= 2\sum_n\left(\sigma^+_n\sigma^-_{n+1} + \sigma^-_n\sigma^+_{n+1}\right), \nonumber\\
V &= 2\sum_n\left(\sigma^+_n\sigma^+_{n+1} + \sigma^-_n\sigma^-_{n+1}\right),
\end{align}
so that $T$ collects the $U(1)$-conserving part of the transverse exchange and $\gamma V = V_\gamma$ is the perturbation~\eqref{eq:Vgamma}.

The two N\'eel product states are degenerate on a finite even ring. We select the required pure phase before applying nondegenerate perturbation theory by adding the regulator $-h\sum_n\epsilon_n\sz_n$, $h>0$, where $\epsilon_n$ is the sign pattern of the N\'eel-up state $|N\rangle$. The calculation is first performed at finite $N$ and $h>0$; after division by $N$ one takes $N\to\infty$ and then $h\to0^+$. None of the paths through fourth order reaches the opposite N\'eel state, and every intermediate state differs from $|N\rangle$ on finitely many sites. A state with $m$ parallel bonds has an excitation energy tending to $2\Dz m$ as $h\to0^+$. Thus, writing $G=Q(E_0-H_0)^{-1}Q$ and $Q=1-|N\rangle\langle N|$ after this limit, each such state contributes the nonsingular resolvent factor $-1/(2\Dz m)$. This regulator supplies the same pure phase as the definition following Eq.~\eqref{eq:Mz_def}.

Both perturbing operators act on bonds: on an antiparallel bond, $T$ interchanges the two spins with amplitude $2$ and $V$ annihilates; on a parallel bond, $V$ reverses both spins with amplitude $2$ and $T$ annihilates. Every bond of $|N\rangle$ is antiparallel, so
\begin{equation}\label{eq:selection}
V|N\rangle = 0, \qquad \langle N|V = 0, \qquad \langle N|T|N\rangle = 0,
\end{equation}
and, since $V$ changes the total magnetization by $\pm 2$ while $T$ conserves it, every term of the perturbation series with an odd number of $V$ insertions vanishes. Observables are therefore even in $\gamma$, as they must be, and the leading $\gamma$-dependence carries two $V$ insertions, neither of which can act first or last on $|N\rangle$.

The staggered magnetization is conveniently written through the density of flipped sites. With $\epsilon_n = \pm 1$ the N\'eel sign pattern of $|N\rangle$ and $P_n = (1 - \epsilon_n\sz_n)/2$ the projector onto a flipped spin at site $n$,
\begin{equation}\label{eq:deficit}
\Mz = 1 - \frac{1}{N}\langle \hat D\rangle, \qquad \hat D = 2\sum_n P_n,
\end{equation}
where $N$ is the number of sites and $\hat D$ is diagonal in the $\sz$ basis, with $\hat D|N\rangle = 0$. With intermediate normalization $\langle N|\psi\rangle = 1$ and $E_1 = \langle N|W|N\rangle = 0$ for $W = T + \gamma V$, the corrections to the ground state read $|\psi_1\rangle = GW|N\rangle$, $|\psi_2\rangle = GWGW|N\rangle$, $|\psi_3\rangle = GWGWGW|N\rangle - E_2 G^2W|N\rangle$ with $E_2 = \langle N|WGW|N\rangle$, and
\begin{align}\label{eq:D_expansion}
\langle\hat D\rangle_{\rm norm} = \frac{1}{\mathcal N}\left[
\langle\psi_1|\hat D|\psi_1\rangle
+ 2\operatorname{Re}\langle\psi_1|\hat D|\psi_2\rangle\right. \nonumber\\
\left.+ \langle\psi_2|\hat D|\psi_2\rangle
+ 2\operatorname{Re}\langle\psi_1|\hat D|\psi_3\rangle + \cdots\right],
\end{align}
where $\mathcal N=\langle\psi|\psi\rangle$. The normalization denominator does not contribute to the term with two $V$ insertions at fourth order in $W$. To see this, work at the finite $N$ specified above. The numerator begins at second order because $\hat D|N\rangle=0$; its $\gamma^2$ part begins at fourth order. The $\gamma^0$ part of $\mathcal N-1$ begins at second order, while its $\gamma^2$ part begins at fourth order. Multiplying either part of $\mathcal N-1$ by the complementary numerator therefore produces a $\gamma^2$ term only at sixth order. Thus $\mathcal N$ may be replaced by unity for the fourth-order coefficient calculated below. That coefficient is proportional to $N$, so division by $N$ gives a finite bulk limit.

\subsection{Zeroth order in \texorpdfstring{$\gamma$}{gamma}}\label{app:pt-warmup}

Acting on $|N\rangle$, $T$ interchanges the spins of any one of the $N$ bonds. The resulting states $|b_i\rangle$ (spins at sites $i, i+1$ interchanged, i.e.\ flipped) have $m = 2$ parallel bonds, so
\begin{equation}\label{eq:psi1}
|\psi_1\rangle\Big|_{\gamma^0} = -\frac{1}{4\Dz}\cdot 2\sum_i |b_i\rangle = -\frac{1}{2\Dz}\sum_i |b_i\rangle .
\end{equation}
Each $|b_i\rangle$ has two flipped sites, $\langle b_i|\hat D|b_i\rangle = 4$, and the states are orthonormal, so $\langle\psi_1|\hat D|\psi_1\rangle = N/\Dz^{2}$ and
\begin{equation}\label{eq:M0_strong}
\Mz(0,\Dz) = 1 - \frac{1}{\Dz^2} + O(\Dz^{-4}),
\end{equation}
which reproduces the Ising expansion of Baxter's formula, $\Mz(0) = 1 - 4x^2 + O(x^4)$ with $x = \Dz - \sqrt{\Dz^2-1}$. This fixes all conventions.

\subsection{Order \texorpdfstring{$\gamma^2$}{gamma squared}}\label{app:pt-gamma2}

By the selection rules~\eqref{eq:selection}, the terms of order $\gamma^2$ in Eq.~\eqref{eq:D_expansion} at fourth order in $W$ are two:
\begin{equation}\label{eq:two_terms}
\langle\hat D\rangle_{\rm norm}\Big|_{\gamma^2} = \gamma^2\left[\langle\chi|\hat D|\chi\rangle
+ 2\langle\psi_1|\hat D|\varphi\rangle\right],
\end{equation}
with $|\chi\rangle = GVGT|N\rangle$ the $\gamma$-linear part of $|\psi_2\rangle$ and $|\varphi\rangle = GVGVGT|N\rangle$ the $\gamma^2$ part of $|\psi_3\rangle$. The energy term in $|\psi_3\rangle$ has no $\gamma$ dependence because $V|N\rangle=\langle N|V=0$. All lower orders vanish: two $V$ insertions at third order would force one of them to be adjacent to $|N\rangle$ on the ket or bra side.

\paragraph*{The state $|\chi\rangle$.} A state $|b_i\rangle$ contains two parallel bonds, $(i-1,i)$ and $(i+1,i+2)$, and $V$ acting there reverses the corresponding pair. In either case the result is a state with two flipped spins on the same sublattice at distance two --- sites $\{i-1,i+1\}$ or $\{i,i+2\}$ --- forming a string of four parallel bonds, $m = 4$. Denote by $|F_j\rangle$ the state with flipped sites $\{j, j+2\}$. Each $|F_j\rangle$ is reached from two intermediate states, $|b_j\rangle$ and $|b_{j+1}\rangle$, and collecting the amplitudes [$2$ from $T$, $2$ from $V$, and resolvents $-1/(4\Dz)$, $-1/(8\Dz)$]:
\begin{align}\label{eq:chi}
|\chi\rangle &= \left(-\frac{1}{8\Dz}\right)(2)\left(-\frac{1}{4\Dz}\right)(2)\cdot 2\sum_j |F_j\rangle \nonumber\\
&= \frac{1}{4\Dz^2}\sum_j |F_j\rangle .
\end{align}
Each $|F_j\rangle$ has two flipped sites, so $\langle F_j|\hat D|F_j\rangle = 4$ and
\begin{equation}\label{eq:term_a}
\langle\chi|\hat D|\chi\rangle = \frac{N}{4\Dz^4}.
\end{equation}

\paragraph*{The state $|\varphi\rangle$.} A state $|F_j\rangle$ contains four parallel bonds. Acting with $V$ on the two interior bonds, $(j,j+1)$ and $(j+1,j+2)$, reverses a pair adjacent to one of the flipped sites and returns a two-flip state: $\{j,j+2\}\triangle\{j,j+1\} = \{j+1,j+2\}$ and $\{j,j+2\}\triangle\{j+1,j+2\} = \{j,j+1\}$, i.e.\ the swap states $|b_{j+1}\rangle$ and $|b_j\rangle$. Acting on the two boundary bonds produces flipped pairs at distance three ($m = 4$), which are orthogonal to $|\psi_1\rangle$ and do not contribute to Eq.~\eqref{eq:two_terms}. Keeping the swap channel, each $|b_i\rangle$ is reached from $|F_{i-1}\rangle$ and $|F_i\rangle$, so its amplitude in $|\varphi\rangle$ is
\begin{equation}\label{eq:phi_amp}
\frac{1}{4\Dz^2}\times 2 \times\left(-\frac{1}{4\Dz}\right)\times 2 = -\frac{1}{4\Dz^3},
\end{equation}
and, with $\langle b_i|\hat D|b_i\rangle = 4$ and the amplitude $-1/(2\Dz)$ of $|b_i\rangle$ in $|\psi_1\rangle$,
\begin{equation}\label{eq:term_b}
2\langle\psi_1|\hat D|\varphi\rangle = 2 N \left(-\frac{1}{2\Dz}\right)(4)\left(-\frac{1}{4\Dz^3}\right) = \frac{N}{\Dz^4}.
\end{equation}

\subsection{Result and interpretation}\label{app:pt-result}

Combining Eqs.~\eqref{eq:deficit}, \eqref{eq:two_terms}, \eqref{eq:term_a} and~\eqref{eq:term_b},
\begin{equation}\label{eq:pt_final}
\frac{\Mz(\gamma,\Dz)}{\Mz(0,\Dz)} = 1 - \frac{5}{4}\frac{\gamma^2}{\Dz^4} + O\left(\frac{\gamma^2}{\Dz^{6}},\gamma^4\right),
\end{equation}
which is Eq.~\eqref{eq:ptclaim}, in agreement with the leading term of the large-$\Dz$ expansion~\eqref{eq:largeDz_Delta} of the closed form; the division by $\Mz(0)$ modifies the result only at $O(\Dz^{-6})$. Equation~\eqref{eq:term_a} is the diagonal weight of the distance-two flipped pairs created by $VT$, while Eq.~\eqref{eq:term_b} is the interference between $T|N\rangle$ and the swap component regenerated by $VVT|N\rangle$. Both matrix elements are positive contributions to the flipped-site density, so their sum lowers the staggered moment. In the closed form the same coefficient comes from the $n=1$ term of the Lambert sum alone (Sec.~\ref{sec:asymp}).

\bibliography{refs}

\end{document}